# Monolithic AlGaAs second-harmonic nanoantennas

**V. F. Gili[1], L. Carletti[2], A. Locatelli[2], D. Rocco[2], M. Finazzi[3], L. Ghirardini[3], I. Favero[1], C. Gomez[4], A. Lemaître[4], M. Celebrano[3], C. De Angelis[2], G. Leo[1,*]**

[1]*Matériaux et Phénomènes Quantiques, Université Paris Diderot - Sorbonne Paris Cité,*
*10 rue A. Domon et L. Duquet, 75013 Paris, France*
[2]*Dept. of Information Engineering, University of Brescia, Via Branze 38, 25123 Brescia, Italy*
[3]*Dept. of Physics, Politecnico di Milano, Piazza Leonardo Da Vinci 32, 20133 Milano, Italy*
[4]*Laboratoire de Photonique et Nanostructures, CNRS-UPR20, Route de Nozay, 91460 Marcoussis, France*
*\*giuseppe.leo@univ-paris-diderot.fr*

We demonstrate monolithic aluminum gallium arsenide (AlGaAs) optical anoantennas. Using a selective oxidation technique, we fabricate such epitaxial semiconductor nanoparticles on an aluminum oxide substrate. Second harmonic generation from an AlGaAs nanocylinder of height h=400 nm and varying radius pumped      with femtosecond pulses delivered at 1554-nm wavelength has been measured, revealing a peak conversion efficiency exceeding $10^{-5}$ for nanocylinders with an otpimized geometry.

## I. Introduction

Metal-less nanophotonics has recently raised an increasing interest because the optical response of high permittivity dielectric nanoparticles in a low refractive index background exhibits negligible dissipative losses and strong magnetic multipole resonances [1–3] in the visible and near-infrared bands of the electromagnetic spectrum. These optical properties can benefit many applications such as directional scattering and emission [4–7], surface enhanced vibrational spectroscopy [8,9], and flat metasurfaces for phase-front engineering [10–12]. Compared to metallic nanoparticles, where the electric field is strongly confined close to the metal surface, the electric field of the resonant modes in dielectric nanoparticles penetrates deep inside their volume, strongly enhancing intracavity light-matter interactions. This fact encourages the use of all-dielectric nanoantennas for strengthening the nonlinear optical response, and some promising results have been achieved in the last two years in the framework of semiconductor technology. In particular, third-harmonic generation (THG) in silicon-on-insulator (SOI) nanoantennas has been investigated [13–15]. By using an array pattern of silicon nanostructures, a THG conversion efficiency higher than $10^{-6}$ was demonstrated, which is the highest value reported for THG on nanoscale thin films using comparable pump energies [15]. Another experimental demonstration showed an increase of two-photon absorption (TPA) by almost two orders of magnitude in hydrogenated amorphous silicon nanodisks with respect to an unstructured silicon film [16]. Although these results clearly illustrate the potential of all-dielectric nanoparticles for nonlinear nanophotonic applications, they are limited by the inherent constraints of the SOI platform in which they have been obtained.

Let us recall that the SOI technology, which is well known for large-scale production of electronic devices, is also becoming a standard for integrated photonics [17]. Its building block is a planar shallow waveguide which strongly confines light thanks to the high refractive index contrast between a silicon (Si) core and a silicon-dioxide ($SiO_2$) substrate ($n_{Si} \approx 3.45$ and $n_{SiO2} \approx 1.44$ at a wavelength around 1.5 µm). Despite the spectacular achievements of silicon technology, the perspective of monolithic SOI photonics seems hardly viable because of a few hard facts: 1) silicon is not a laser material due to its indirect energy gap; 2) it is affected by TPA at optical telecom wavelengths due to its gap of 1.1 eV; 3) it does not exhibit quadratic optical nonlinearity, because of its centro-symmetric crystal structure. Owing to the latter two reasons, only $\chi^{(3)}$ nonlinear effects have been observed so far in dielectric nanoantennas. In order to overcome on the one hand these limitations, and on the other hand the technological burden of the wafer bonding step associated to the recently reported hybridization of AlGaAs on the SOI system [18], here we propose an AlGaAs-based monolithic platform for nonlinear nanophotonics.

III-V semiconductor $Al_xGa_{1-x}As$ is a popular material for integrated nonlinear optics [19] because, besides a few properties that it shares with Si (a similar refractive index in the near infrared, a $\chi^{(3)}$ nonlinear index, a good thermal conductivity), it also exhibits some key comparative advantages: 1) a direct gap which increases with Al molar fraction $x$ ($E_g = 1.424 + 1.266x + 0.26x^2$ eV, for $x < 0.45$) and enables TPA-free operation at 1.55 µm for $x \geq 0.18$; 2) a huge non-resonant quadratic susceptibility ($d_{14} \approx 110$ pm/V for GaAs in the near infrared); 3) a wide transparency window in the mid-infrared (up to 17 µm); and 4) a mature heterostructure laser technology. However, a full development of an AlGaAs photonic platform has been hindered so far by the difficulty of fabricating monolithic shallow nanowires as in the SOI system, although the selective oxidation of AlGaAs epitaxial layers was discovered in 1990 and gives rise to an aluminium oxide (AlOx) with optical and electrical properties similar to $SiO_2$ [20]. Let us point out that while the use of AlOx layers thinner than 100 nm is standard in VCSEL technology and has resulted in the demonstration of an optical parametric oscillator in an AlGaAs waveguide [21], the fabrication of devices with high-quality micron-thick AlOx optical substrates has not been reported to date.

In this paper we first describe the fabrication of $\chi^{(2)}$ AlGaAs-on-AlOx monolithic nanostructures on a GaAs wafer, which we carried out by an appropriate design of the epitaxial structure and a careful control of the oxidation parameters (Section 2). Then, based on numerical modelling, we predict the occurrence of second harmonic generation (SHG) in $Al_{0.18}Ga_{0.82}As$ nanocylinders on AlOx, with a conversion efficiency up to $7\times10^{-5}$ for a 1.6 GW/cm$^2$ pump in the optical telecom wavelength range (Section 3). Finally, in Section 4 we report and discuss the first experimental demonstration of nonlinear response AlGaAs-on-AlOx nanoantennas, in fair agreement with such predictions.

## II. Fabrication

The selective oxidation of AlGaAs optical substrates requires some care, in order to avoid that the oxidation-induced AlGaAs contraction results in defects at the interfaces between the non-stoichiometric amorphous AlOx and the adjacent crystal lattices.

Our samples were grown by molecular-beam-epitaxy on [100] non-intentionally doped GaAs wafer, with a 400 nm layer of $Al_{0.18}Ga_{0.82}As$ on top of an aluminum-rich substrate, to be oxidized at a later stage. In order to improve the eventual adhesion between AlOx and the adjacent crystalline layers, such substrate consists of a 1-μm-thick aluminium-rich AlGaAs layer sandwiched between two transition regions with varying aluminium molar fraction. In order to obtain an array of nanodisks as shown in Fig. 1a, we patterned circles with radii between 180 nm and 220 nm, and equally spaced by 3 μm, with a scanning electron microscope (SEM) Zeiss lithography system. Then our samples were dry etched with non-selective ICP- RIE (Sentech SI500) with $SiCl_4$:Ar chemical treatment. The etching depth of 400 nm, controlled by laser interferometer, defined the nano-cylinders and revealed the AlAs. Then the etched sample was oxidized at 390°C for 30 minutes in an oven equipped with in situ optical monitoring, under a precisely controlled water vapor flow with $N_2$:$H_2$ gas carrier. After oxidation, each $Al_{0.18}Ga_{0.82}As$ nanocylinder lies upon a uniform AlOx substrate (see Fig. 1b), whose low refractive index (n≈1.6) enables sub-wavelength optical confinement in the nanocavity by total internal reflection.

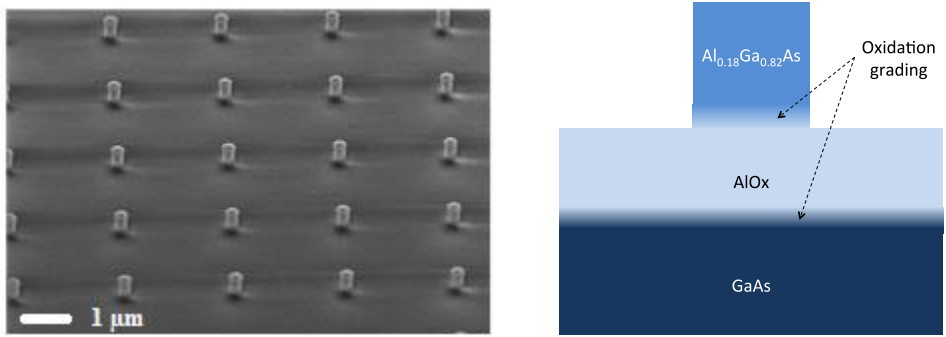

**Fig. 1**: Monolithic AlGaAs-on-AlOx nanoantennas: a) scanning-electron-microscope picture of a part of the array; b) schematics of a single nanoantenna.

## III. Numerical modeling

The linear and nonlinear optical response of the disk nanoantennas on an aluminum-oxide (AlOx) substrate are modelled by using frequency-domain finite element simulations [22]. The scattering efficiency spectra as a function of disk radius are shown in Fig. 2(a). We can observe that the three resonances that appear in this wavelength range red-shift while the radius is increased. In Fig. 2(b) the scattering efficiency spectra of a disk with a radius of 225 nm is obtained by extracting the data along the dashed line on Fig. 2(a). In order to gain a deeper physical insight on the behavior of the nanoantenna, the optical response is expanded using a multipole decomposition. Because of the presence of the substrate, the coefficients associated to the various multipole contributions are obtained via integration of the dielectric currents induced in the AlGaAs disk by the incident field [23,24]. For simplicity, the cross-sections associated with each multipole are calculated as if it was radiating into air, thus the total cross-section is slightly higher than the sum over multipole contributions. As it can be seen from Fig. 2(b), the resonance at a wavelength of 1660 nm is due to a magnetic dipole (MD) mode over an electric dipole (ED) contribution. This is confirmed by the calculation of the electric field in the cylinder at this wavelength, which shows that the electric field loop that characterizes the MD mode partially leaks into the AlOx substrate [25].

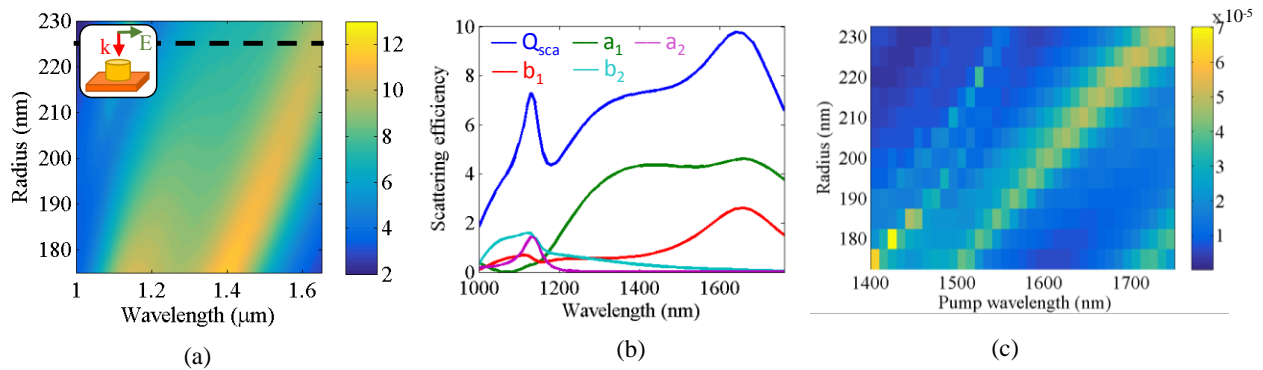

**Fig. 2.** (a) Scattering efficiency spectra as a function of the nanoantenna radius and for a fixed nanoantenna height of 400 nm. The dashed line represents a antenna with radius r = 225 nm. The schematics of the illumination is shown in the inset. Light is at normal incidence on the nanoantenna with electric field E linearly polarized parallel to the substrate plane. (b) Scattering efficiency ($Q_{sca}$) spectra for a disk with r = 225 nm, with the multipole expansion terms of the optical response. The coefficients $a_1$, $b_1$, $a_2$, and $b_2$ are due to electric dipole, magnetic dipole, electric quadrupole, and magnetic quadrupole contributions, respectively. (c) Electric field enhancement map in the incidence plane at the MD resonance. Cones indicate the direction of electric field vector.

The second-order nonlinear optical response associated with the SHG is modeled using the nonlinear polarization induced by $\chi^{(2)}$ as a source in COMSOL simulations [22]. Since AlGaAs has a zinc blende crystalline structure, the only non-vanishing terms of the nonlinear susceptibility tensor are the $d_{ijk}$ terms with $i{\neq}j{\neq}k$, which, in the calculations, have a value of 100 pm/V. The conversion efficiency is defined as

$$\eta_{SHG} = \frac{\int\limits_{A} \vec{S}_{SH} \cdot \hat{n}\, da}{I_0 \times \pi r^2}$$

where $\vec{S}_{SH}$ is the Poynting vector of the SH field, $\hat{n}$ is the unit vector normal to a surface A enclosing the antenna, and $I_0$ is the incident field intensity. ($I_0$ = 1 GW/cm² in the simulations). Fig. 2(c) shows the calculated SH conversion efficiency as a function of the pump wavelength and for varying disk radius. We numerically find a SH conversion efficiency up to $6{\times}10^{-5}$ for a pump wavelength between 1500 nm and 1700 nm while the disk radius is varying between 175 nm and 230 nm. The SH conversion efficiency is smaller than the value reported in Ref. 22 due to the field leakage effect caused by the substrate that hampers the strength of the MD resonance. The SHG peak at a pump wavelength in the range between 1500 nm and 1700 nm while the radius is varied originates from the good spatial and spectral overlap between the nonlinear currents induced by the pump and the resonant mode of the disk at the emission wavelength [22].

## IV. Experimental results and discussion

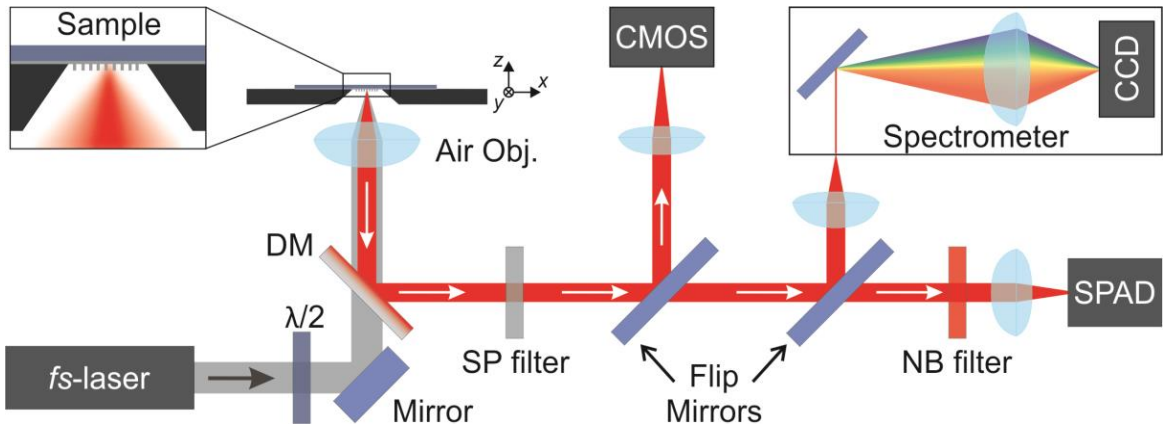

**Figure 3**: Experimental set-up for the nonlinear spectroscopy of individual nanoantennas. Fs-laser: 1554 nm Er-doped fiber laser delivering 150-fs pulses (light grey beam). λ/2: half-wave plate; Air Obj.: 0.85-NA air objective; DM: dichroic mirror; SP & NB filters: short-pass and narrow-band filter. CMOS: imaging camera; SPAD: single-photon avalanche detector; CCD: spectrometer cooled camera. The sample is mounted facing downwards as depicted in the inset.

To experimentally investigate the nonlinear properties of the fabricated nanopillars and validate the theoretical description drawn above, we employed a nonlinear microscopy setup in epi-reflection configuration as sketched in Figure 3. The excitation light is delivered by a linearly polarized (extinction ratio ≈ 100:1) ultrafast Erbium-doped fiber laser centred at 1554 nm laser (150 fs pulse length). The light polarization is finely-tuned by means of a λ/2 plate before entering the objective. The excitation beam is tightly focused onto the sample through a high-NA objective (Nikon, NA = 0.85) to allow the interrogation of individual pillars inside the array under study. The light emitted by the sample, collected through the same optical path, is reflected by a polarization-mantaining dichroic mirror (DMLP1180, Thorlabs Inc.) to reject the excitation light. To further clean-up the nonlinear signal we inserted a short-pass filter (SP01-785RU-25, Semrock - IDEX Health & Science, LLC.) inside the detection line, which selects solely the visible portion of the electromagnetic spectrum. The spectral components of the nonlinear emitted light are investigated using a visible/near-infrared grating-based spectrometer (Shemrock 303i, Andor Technology). An alternative detection path for the the emitted SHG consists in a narrow-band filter at 775 nm (NB filter, 40 nm bandwidth, Samrock) followed by a Single Photon Avalanche Detector (SPAD). SHG maps can be eventually recorded by simoultaneously scaning the hollow piezo (P-517.3CL, Phyisik Instrumente) on which the sample is placed.

SHG signals were collected using the detection path including the SPAD from an array of nanocylinders with radius varying from 175 to 225 nm. The dependence of the experimentally detected SHG on the radius size of the nanocylindrs is reported in Fig. 4a (hollow diamonds). A multi-Lorenzian fit of the curve identifying three features.

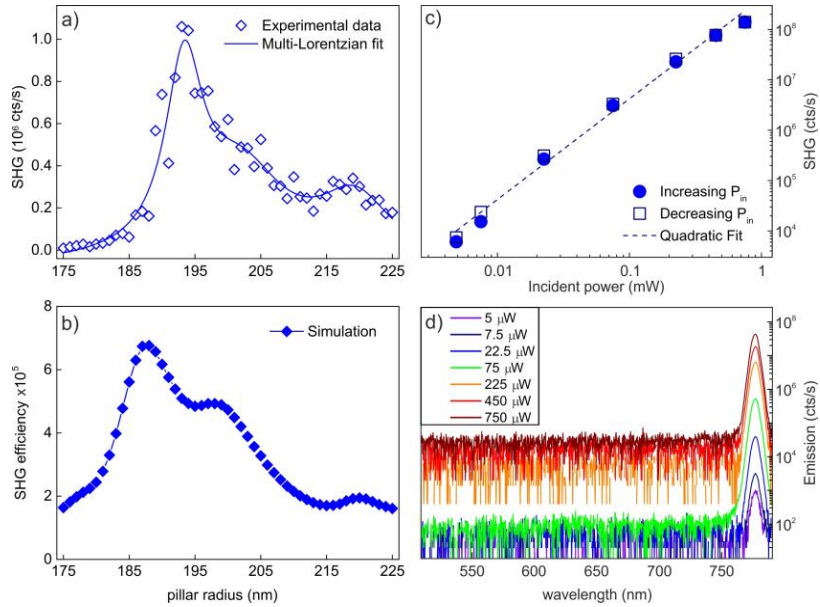

**Figure 4**: a) Measured SHG intensity rate as a function of the nanoantenna radius (hollow diamonds). As a guide to the eye a multi-Gaussian fit featuring three distinct peaks. b) Nunmerical calculations of the SHG efficiency on nanoantennas featuring the same geometrical parameters as the fabricated/measured ones. c) Power curve in Log/Log scale. SHG intensity as a function of the pump intensity for nanoantenna with 193 nm radius. d) Emission spectra as a function of the incident power.

The calculated radius-dependent behavior of the conversion efficiency reported in Fig. 4b singles out three resonances for the SHG nanocavity modes excited at three specific radii size, in striking agreement with the experimental values of the overall SHG collected signal. The small shift of about 5 nm that can be observed on the main peak of the curve is attributed to fabrication tolerances between the nominal and the actual size of the nanocylinders.

We have measured the SHG intensity emitted by the most efficient nanoantenna (193-nm radius) for a illumination power ranging from 5 to 750 μW. Figure 4c reports the power-dependent curves of the SHG signal, retrieved from the signal recorded at the SPAD side. The full cyrcles identifies the SHG photon counts recorded while increasing the incident power level, whereas the hollow squares indicates the photon counts while decrasing the input power. The excellent overlap between the two curves indicates that no degradation occurs in this power interval, while the quadratic fit to the curves (dashed line) confirms the second order origin of this process. Figure 4c reveals a peak count rate of about $1.4 \times 10^8$ photons per second, which corresponds to an effective SHG photon flux of about $3.2 \times 10^{10}$ per second leading to a peak conversion efficiency of about $1.1 \times 10^{-5}$. These crucial figures are obtained by accounting for the SPAD filling-factor (0.34), the optical system throughput (0.13), and the objective collection efficiency (0.1). The latter stems from the ratio between the collected and the total SHG power, as estimated by the calculated radiation pattern. Noteworthy, the peak efficiency evaluated with our calculations is of the same order of magnitude as the one estimated from measurements.

By flipping the mirror in front of the SPAD it is possible to couple the collected light to the spectrometer and measure the emission spectrum in the visible/near-infrared range. Figure 4d shows the emission spectra recorded at the same power levels employed for the intensity curve in Fig. 4c. Regardless of the pump power these nanocylinders display neither any sizeable multiphoton photoluminescence nor third harmonic generation.

**V. Conclusions**

We have demonstrated monolithic AlGaAs optical nanoantennas fabricated on an AlOx substrate. Using optimized selective oxidation technique a high-quality micron-thick AlOx optical substrates for device fabrication was obtained. These nanoantennas have been probed for SHG driven by a magnetic dipole resonance at the wavelength of 1550 nm. The experimental SHG spectra and conversion efficiency (about $10^{-5}$ at 1.6 GW/cm$^2$ pump intensity) are in excellent agreement with the numerical simulations. These results show the huge potential of AlGaAs nanoantennas for nonlinear nanophotonic applications.

**Acknowledgments**

This work was carried out in the framework of the Erasmus Mundus NANOPHI project, contract number 2013 5659/002-001, and of the NanoSpectroscopy COST Action MP1302. V. F. Gili thanks Mehdi Hamoumi and William Hease for fabrication hints, and acknowledges financial support from the "Double Culture" PhD Program of Sorbonne Paris Cité. The authors acknowledge financial support from CARIPLO Foundation (project "SHAPES - Second-HArmonic Plasmon-Enhanced Sensing" ref. 2013-0736), and thank P. Biagioni, G. Marino, G. Pellegrini, and A. Zayats for fruitful discussions.

---

**References and links**

1. A. I. Kuznetsov, A. E. Miroshnichenko, Y. H. Fu, J. Zhang, and B. Luk'yanchuk, "Magnetic light," Sci. Rep. **2**, 1–6 (2012).

2.  A. B. Evlyukhin, S. M. Novikov, U. Zywietz, R. L. Eriksen, C. Reinhardt, S. I. Bozhevolnyi, and B. N. Chichkov, "Demonstration of magnetic dipole resonances of dielectric nanospheres in the visible region," Nano Lett. **12**, 3749–3755 (2012).

3.  S. Person, M. Jain, Z. Lapin, J. J. Sáenz, G. Wicks, and L. Novotny, "Demonstration of zero optical backscattering from single nanoparticles," Nano Lett. **13**, 1806–1809 (2013).

4.  W. Liu, J. Zhang, B. Lei, H. Ma, W. Xie, and H. Hu, "Ultra-directional forward scattering by individual core-shell nanoparticles," Opt. Express **22**, 16178 (2014).

5.  B. Rolly, B. Stout, and N. Bonod, "Boosting the directivity of optical antennas with magnetic and electric dipolar resonant particles," Opt. Express **20**, 20376 (2012).

6.  Y. H. Fu, A. I. Kuznetsov, A. E. Miroshnichenko, Y. F. Yu, and B. Luk'yanchuk, "Directional visible light scattering by silicon nanoparticles.," Nat. Commun. **4**, 1527 (2013).

7.  M. Celebrano, M. Baselli, M. Bollani, J. Frigerio, A. Bahgat Shehata, A. Della Frera, A. Tosi, A. Farina, F. Pezzoli, J. Osmond, X. Wu, B. Hecht, R. Sordan, D. Chrastina, G. Isella, L. Duò, M. Finazzi, and P. Biagioni, "Emission Engineering in Germanium Nanoresonators," ACS Photonics 2, 53-59 (2014).

8.  N. Bontempi, L. Carletti, C. De Angelis, and I. Alessandri, "Plasmon-free SERS detection of environmental $CO_2$ on $TiO_2$ surfaces," Nanoscale **8**, 3226–3231 (2016).

9.  M. Caldarola, P. Albella, E. Cortés, M. Rahmani, T. Roschuk, G. Grinblat, R. F. Oulton, A. V. Bragas, and S. A. Maier, "Non-plasmonic nanoantennas for surface enhanced spectroscopies with ultra-low heat conversion," Nat. Commun. **6**, 7915 (2015).

10. I. Staude, A. E. Miroshnichenko, M. Decker, N. T. Fofang, S. Liu, E. Gonzales, J. Dominguez, T. S. Luk, D. N. Neshev, I. Brener, and Y. Kivshar, "Tailoring directional scattering through magnetic and electric resonances in subwavelength silicon nanodisks," ACS Nano **7**, 7824–7832 (2013).

11. M. Decker, I. Staude, M. Falkner, J. Dominguez, D. N. Neshev, I. Brener, T. Pertsch, and Y. S. Kivshar, "High-Efficiency Dielectric Huygens' Surfaces," Adv. Opt. Mater. **3**, 813–820 (2015).

12. A. Arbabi, Y. Horie, M. Bagheri, and A. Faraon, "Dielectric metasurfaces for complete control of phase and polarization with subwavelength spatial resolution and high transmission," Nat. Nanotechnology 1–8 (2015).

13. M. R. Shcherbakov, D. N. Neshev, B. Hopkins, A. S. Shorokhov, I. Staude, E. V Melik-Gaykazyan, M. Decker, A. A. Ezhov, A. E. Miroshnichenko, I. Brener, A. A. Fedyanin, and Y. S. Kivshar, "Enhanced Third-Harmonic Generation in Silicon Nanoparticles Driven by Magnetic Response," Nano Lett. **14**, 6488–6492 (2014).

14. M. R. Shcherbakov, A. S. Shorokhov, D. N. Neshev, B. Hopkins, I. Staude, E. V. Melik-Gaykazyan, A. A. Ezhov, A. E. Miroshnichenko, I. Brener, A. A. Fedyanin, and Y. S. Kivshar, "Nonlinear Interference and Tailorable Third-Harmonic Generation from Dielectric Oligomers," ACS Photonics **2**, 578–582 (2015).

15. Y. Yang, W. Wang, A. Boulesbaa, I. I. Kravchenko, D. P. Briggs, A. Puretzky, D. Geohegan, and J. Valentine, "Nonlinear Fano-Resonant Dielectric Metasurfaces," Nano Lett. **15**, 7388–7393 (2015).

16. M. R. Shcherbakov, P. P. Vabishchevich, A. S. Shorokhov, K. E. Chong, D.-Y. Choi, I. Staude, A. E. Miroshnichenko, D. N. Neshev, A. A. Fedyanin, and Y. S. Kivshar, "Ultrafast All-Optical Switching with Magnetic Resonances in Nonlinear Dielectric Nanostructures," Nano Lett. **15**, 6985–6990 (2015).

17. Focus issue on silicon photonics, "Simply silicon," Nat. Photonics **4**, 491–544 (2010).

18. M. Pu, L. Ottaviano, E. Semenova, K. Yvind, "AlGaAs-On-Insulator Nonlinear Photonics," arXiv 21 (2015).

19. A. S. Helmy, P. Abolghasem, J. Stewart Aitchison, B. J. Bijlani, J. Han, B. M. Holmes, D. C. Hutchings, U. Younis, and S. J. Wagner, "Recent advances in phase matching of second-order nonlinearities in monolithic semiconductor waveguides," Laser Photonics Rev. **5**, 272–286 (2011).

20. J. M. Dallesasse, N. Holonyak, A. R. Sugg, T. A. Richard, N. El-Zein, "Hydrolyzation oxidation of $Al_xGa_{1-x}As$-AlAs-GaAs quantum well heterostructures and superlattices," Appl. Phys. Lett. **57**, 2844–2846 (1990).

21. M. Savanier, C. Ozanam, L. Lanco, X. Lafosse, A. Andronico, I. Favero, S. Ducci, and G. Leo, "Near-infrared optical parametric oscillator in a III-V semiconductor waveguide," Appl. Phys. Lett. **103**, 261105 (2013).

22. L. Carletti, A. Locatelli, O. Stepanenko, G. Leo, and C. De Angelis, "Enhanced second-harmonic generation from magnetic resonance in AlGaAs nanoantennas," Opt. Express **23**, 26544–26550 (2015).

23. P. Grahn, A. Shevchenko, and M. Kaivola, "Electromagnetic multipole theory for optical nanomaterials," New J. Phys. **14**, 093033 (2012).

24. R. M. Bakker, D. Permyakov, Y. F. Yu, D. Markovich, R. Paniagua-Domínguez, L. Gonzaga, A. Samusev, Y. S. Kivshar, B. Luk`yanchuk, and A. I. Kuznetsov, "Magnetic and Electric Hotspots with Silicon Nanodimers," Nano Lett. **15**, 2137–2142 (2015).

25. J. Van De Groep and A. Polman, "Designing dielectric resonators on substrates : Combining magnetic and electric resonances," Opt. Express **21**, 1253–1257 (2013).